\documentclass[11pt,ams]{article}
\usepackage{graphicx}
\usepackage{amsmath}

\begin{document}

\date{}
\title{\textbf{Soft breaking of BRST invariance for  introducing  non-perturbative infrared effects in a local and
renormalizable way}}
\author{\textbf{L. Baulieu$^{ab}$\thanks{%
baulieu@lpthe.jussieu.fr}}\,, \textbf{S.P. Sorella}$^{c}$\thanks{%
sorella@uerj.br}\; \\
\\
\textit{$^a$ {\small Theoretical Division CERN}}\thanks{CH-1211 Gen{\`e}ve, 23, Switzerland}\\[3mm]
\textit{$^b$ {\small LPTHE, CNRS and Universit{\'e} Pierre et Marie Curie}}\thanks{4 place Jussieu, F-75252 Paris Cedex 05, France} \\[3mm]
\textit{$^{c}${\small {UERJ $-$ Universidade do Estado do Rio de
Janeiro}}}\\
\textit{{\small {Instituto de F\'{\i }sica $-$ Departamento de F\'{\i }sica
Te\'{o}rica}}}\\
\textit{{\small {Rua S{\~a}o Francisco Xavier 524, 20550-013 Maracan{\~a},
Rio de Janeiro, Brasil}}}}
\maketitle

\begin{abstract}
The possibility of introducing non-perturbative infrared effects
leading to a modification of the long distance behavior of gauge
theories through a soft breaking of the $BRST$ invariance is
investigated. The method reproduces the Gribov-Zwanziger action
describing the restriction of the domain of integration in the
Feynman path integral to the Gribov region and a   model for the
dynamical quark mass generation is presented. The soft symmetry
breaking relies  on the introduction of $BRST$ doublets and
massive physical parameters,  which  allow one to distinguish the
infrared region  from the  ultraviolet one, within the same
theory.

\end{abstract}

\vspace{0.5cm}

\newpage

\def\cor{$\clubsuit$}

\section{Introduction}

The task of introducing non-perturbative effects is one of the
most difficult challenges in quantum field theory. In this letter
we  propose to introduce non-perturbative infrared effects in a
way that preserves the basic properties of continuum field theory,
namely: locality and renormalizability. The idea relies on the
possibility of introducing in a controllable way local terms which
give rise to a soft breaking of the $BRST$ symmetry, meaning that
the dimension of these breaking terms is smaller than the
space-time dimension. Such soft  terms can be neglected in the
deep ultraviolet region, where one recovers the notion of exact
$BRST$ invariance. However, in the low energy infrared region, the
soft  breaking terms cannot be neglected, and  produce a
quantitative modification of the large distance behavior of the
theory. Furthermore, these terms are introduced in a way that
preserves the nilpotency of the $BRST$ operator and doesn't modify
its cohomology. As a consequence, the space of the local
observables of the theory, identified with the cohomology classes
of the $BRST$ operator in the space of local field polynomials,
remains unchanged. In other words, assuming that the relevant
correlation functions of the theory are of the type $ \langle
O_1(x_1)....O_n(x_n) \rangle $ with $O_1(x_1),...,O_n(x_n)$ field
polynomials belonging to the cohomology of the $BRST$ operator, we
attempt at introducing nonperturbative effects which might modify
the infrared behavior of $ \langle O_1(x_1)....O_n(x_n) \rangle $,
while leaving unchanged its behavior in the deep ultraviolet
region. Thus, the same correlation function can display a
different behavior, according to the region which is being
considered. Our aim here is precisely that of keeping the same set
of observables of the starting theory, while modifying the large
distance behavior of their correlation functions.
\newline
\newline
The first step in order to achieve this goal is that of
introducing a suitable set of additional fields, assembled  as
$BRST$ doublets. Such fields do not modify the cohomology of the
$BRST$ operator.  Thus, their introduction  leaves unchanged the
original content of observables of the theory. This follows from
the fact that $BRST$ doublets always give rise to invariant terms
which can be cast in the form of exact $BRST$ variations, thus
yielding vanishing cohomology \cite{Piguet:1995er}. \newline
\newline
Secondly, these $BRST\;$doublets are coupled to a set of
dimensionfull  parameters in such a way that the initial $BRST$
invariance of the theory can only be  softly broken by terms which
can be kept under control at the quantum level. This nontrivial
requirement can be fulfilled thanks to the $BRST$ doublet
structure of the fields coupled to these dimensionful parameters,
which determine suitable linearly broken Ward identities, as well
as generalized Slavnov--Taylor identities, ensuring the
renormalizability  and the stability of the theory in the presence
of the breaking. \newline
\newline
The last step is that of requiring that these parameters are not
treated as free parameters of the theory. Instead, they are
determined in a self-consistent way at the quantum level by
imposing appropriate gap equations, reflecting the possible
non-perturbative condensation of the local operators coupled to
them. This allows one to express them in terms of the original
coupling constant $g$ of the theory, exhibiting the typical
non-perturbative behavior $\sim e^{-\frac{1}{g^{2}}}$. Albeit we
cannot specify in full generality the explicit form of these gap
equations, it is worth underlining that the dimension of the
operators coupled to these parameters is smaller than the
space-time dimensions. The analytic study of the possible
condensation of these operators is simplified as compared to the
case of an operator of maximum dimension. In fact, several
techniques are nowadays at our disposal in order to face the study
of the condensation of local operators of lower dimension:
effective potential, local composite operator technique, the
renormalization group equations, variational principles as the
minimal sensitivity principle, the fastest apparent convergence
criterion, etc. Let us mention, for example, the case of the
condensation of the dimension two mass operator $\left\langle
\left( \frac{1}{2}A^{2}-\alpha \overline{c}c\right) \right\rangle
$ in the Yang--Mills theory expressed in the maximal Abelian
gauge. The condensation of this operator, see \cite{Dudal:2004rx}
and references therein, provides a non-perturbative dynamical mass
for off-diagonal gluons, a feature of great relevance for the
Abelian dominance hypothesis for the dual superconductivity
picture of color confinement.
\newline
\newline
One can certainly  question  the meaning of introducing a soft
breaking of the $BRST$ symmetry, since the latter is known to be
related to physical properties of the theory, such as the
unitarity and the characterization of the physical subspace. The
justification is for models displaying non-perturbative phases,
which require a better knowledge of the infrared sector. This is
the case, for example, of the $QCD$~confining non-abelian gauge
theories. As is well known, the  particle  interpretation of
$QCD$~in terms of partons~-gluons and quarks-~is lost in the low
energy region. Unitarity cannot and shouldn't be defined for the
gluon and quark sector in the infrared region, where confinement
sets in, and gluons and quarks do not correspond anymore to the
physical excitations of the theory. Rather, the physical states
are given by colorless bound states like baryons, mesons and
glueballs. Moreover, the  framework that we introduce here can
also include the description of non-perturbative long distance
effects in supersymmetric as well as in topological theories, both
by taking into account the effect of the Gribov phenomenon and a
soft breaking of the  supersymmetry, as in the Seiberg--Witten
theory \cite{lb}. \newline
\newline
As a last remark, we would like to underline that the soft breaking of the $%
BRST$ symmetry introduced here is meant to be an explicit
breaking, as opposed to the possibility of achieving a spontaneous
symmetry breaking. Indeed, the concept of spontaneous $BRST$
symmetry breaking would imply the existence of Goldstone massless
modes, which would give rise to a rather different framework.
\newline
\newline
The paper is organized as follows. In Sect.2 we give an overview
of the introduction of the $BRST$ doublet fields and of the
related dimensionfull parameters. We point out that the presence
of the soft breaking of the $BRST$ invariance ensures that the
aforementioned dimensionfull parameters are physical parameters of
the theory and not unphysical ones, as it would be the case of
gauge parameters.  An elementary algebraic proof of this property
is given.  Thus, in the infrared domain, the values  of the
correlation functions of the theory will explicitly  depend on
these parameters, which would themselves be related to
$\Lambda_{QCD}$ by generalized gap equations. In Sect.3, we show
that the method yields  the  so-called Gribov-Zwanziger action
which implements the restriction of the domain of integration in
the path integral to the Gribov region, plus a relevant soft term.
In Sect.4 we present a model for the dynamical quark mass
generation.

\section{Soft breaking of the BRST invariance}

We start by considering a quantized action $S_{\mathrm{inv}}(\phi
)$, endowed with a set of fields generically denoted by $\phi$,
with  the following properties:

\begin{itemize}
\item the action  exhibits exact $BRST$ invariance
\begin{equation}
sS_{\mathrm{inv}}(\phi )=0\;,  \label{a1}
\end{equation}
where $s$ denotes the nilpotent $BRST$ operator
\begin{equation}
s^{2}=0\;,  \label{a2}
\end{equation}

\item  $S_{\mathrm{inv}}$ is multiplicatively renormalizable
leading, in particular, to a consistent perturbation theory in the
ultraviolet region.
\end{itemize}
A typical example of such an action is provided by non-abelian
$SU(N)$ Euclidean Yang-Mills theories, namely
\begin{equation}
S_{\mathrm{inv}}=S_{YM}+S_{gf}\,\;,  \label{a3}
\end{equation}
where $S_{YM}$ stands for the Yang--Mills action
\begin{equation}
S_{YM}=\frac{1}{4}\int d^{4}x\,F_{\mu \nu }^{a}F_{\mu \nu }^{a}\;,
\label{a4}
\end{equation}
and $S_{gf}$ is the gauge-fixing term of the Landau gauge
\begin{equation}
S_{gf}=\int d^{4}x\;\left( ib^{a}\partial _{\mu }A_{\mu }^{a}+\overline{c}%
^{a}\partial _{\mu }D_{\mu }^{ab}c^{b}\right) \;,  \label{a5}
\end{equation}
with $b^{a}$ being the Lagrange multiplier field enforcing the
Landau condition, $\partial _{\mu }A_{\mu }^{a}=0$, and $\left(
\overline{c}^{a},c^{b}\right) $ the Faddeev-popov ghosts. As
required, expression $\left( \ref{a3}\right) $ is renormalizable,
while displaying $BRST$ invariance~\cite{Piguet:1995er}.
\newline
\newline
We will now extend this local action for tentatively introducing
non-perturbative infrared effects within the context of local
quantum field theory, which will  lead to a modification of the
long distance behavior of the theory, while leaving unaffected its
ultraviolet behavior.  To do so,
we introduce a set of fields, generically denoted as $\left(
\underline{\alpha },\underline{\beta }\right)$, which transform as
$BRST$ doublets
\begin{eqnarray}
s\underline{\alpha } &=&\underline{\beta }\;,  \nonumber \\
s\underline{\beta } &=&0\;.  \label{a8}
\end{eqnarray}
Such fields give rise to a trivial cohomology, meaning
that any local $BRST$ invariant functional $F(\underline{\alpha },\underline{%
\beta })$ has necessarily the form of an exact $BRST$ term
\begin{equation}
sF(\underline{\alpha },\underline{\beta })=0\;\Rightarrow \;F(\underline{%
\alpha },\underline{\beta })=s\widehat{F}(\underline{\alpha },\underline{%
\beta })\;,  \label{a9}
\end{equation}
for some local $\widehat{F}(\underline{\alpha },\underline{\beta })$.
\newline
\newline
After the introduction of the fields $(\underline{\alpha },\underline{\beta }%
)$ and of a corresponding $BRST$ invariant term describing their propagation,
\begin{equation}
S_{\mathrm{inv}}^{\alpha \beta }(\underline{\alpha },\underline{\beta })=s%
\widehat{S}(\underline{\alpha },\underline{\beta }) \;  \label{din}
\end{equation}
one defines a set of dimensionfull  parameters $\left(
\underline{\sigma }\right) $ and a local term $S_{\sigma
}($\underline{$\sigma $},$\underline{\alpha },\underline{\beta
},\phi )$ which gives rise to a soft breaking of the $BRST$
invariance, as expressed by
\begin{equation}
s\left( S_{\mathrm{inv}}(\phi )+S_{\mathrm{inv}}^{\alpha \beta }(\underline{%
\alpha },\underline{\beta })+S_{\sigma }(\underline{\sigma },\underline{%
\alpha },\underline{\beta },\phi )\right) =sS_{\sigma }=\underline{\sigma }%
\,\Delta (\underline{\alpha },\underline{\beta },\phi )\;.  \label{a10}
\end{equation}
Since $\left( \underline{\sigma }\right) $ are dimensionfull
parameters, the breaking term $\Delta (\underline{\alpha
},\underline{\beta },\phi )$ is an integrated local field
polynomial whose dimension is smaller that the space-time
dimension, \textit{i.e. }it is a soft breaking. As such, it can be
neglected in the deep ultraviolet region. Moreover, taking into
account
that the additional fields $(\underline{\alpha },\underline{\beta })$ form $%
BRST$ doublets, it follows that the ultraviolet behavior of the model is
left unmodified. \newline
\newline
The introduction of the soft parameters $\left( \underline{\sigma
}\right) $ has to be done in a way that preserves the
multiplicative renormalizability of the modified theory. This
physical requirement allows one to carry out calculations in a
consistent way. Furthermore, it  puts severe constraints on the
way  the soft parameters are introduced. The simplest and
consistent way is to couple the additional doublet fields $\left(
\underline{\alpha },\underline{\beta }\right)$ linearly to the
original fields $\phi$, so that the resulting breaking term is a
quadratic term in the fields. As a consequence, the propagator of
the $\phi$-field gets modified in the infrared by the soft
parameters entering the quadratic coupling. In turn, this will
affect the infrared behavior of the correlation functions of the
theory, as can be seen, in particular, from the modifications at
small momentum that are brought to the propagators of partons.
Coupling the fields in that way has important consequences on the
Ward identities of the theory. In fact, a quadratic term means
that the additional fields $\left( \underline{\alpha
},\underline{\beta }\right)$ couple linearly to the original
fields $\phi$. As we shall see in the examples discussed in the
next sections, this feature enables one to write down a suitable
set of linearly broken Ward identities,  controlling the
renormalizability of the theory. In summary, the introduction of
the infrared soft breaking can   be done in a way in which the
resulting action is stable against quantum corrections. This will
ensure that no additional parameters have
to be introduced. \\\\ The presence of the breaking term $\Delta (%
\underline{\alpha },\underline{\beta },\phi )$ enables one to
establish that the dimensionfull parameters $\left(
\underline{\sigma }\right) $ are physical parameters of the
theory, as for instance the Gribov parameter $\gamma$
\cite{Dudal:2007cw,Dudal:2008sp} of the Gribov-Zwanziger action.
In fact, taking the derivative of
both sides of eq.$\left( \ref{a10}\right) $ with respect to $\underline{%
\sigma }$,  one gets
\begin{equation}
s\frac{\partial S_{\sigma }}{\partial \underline{\sigma }}=\Delta \;,
\label{a11}
\end{equation}
from which it is apparent that $\frac{\partial S_{\sigma }}{\partial
\underline{\sigma }}$ cannot be cast in the form of an exact $BRST$
variation, namely
\begin{equation}
\frac{\partial S_{\sigma }}{\partial \underline{\sigma }}\neq s\widehat{%
\Delta }\;,  \label{a12}
\end{equation}
for some local $\widehat{\Delta }$. Equation $\left(
\ref{a12}\right) $ ensures that the parameters $\left(
\underline{\sigma }\right) $ are physical parameters of the
theory. As such, they will influence   the expressions of the
correlation functions of the operators belonging to the cohomology
os $s$, modifying their long distance behavior. We see therefore
that the soft breaking term $\Delta $ plays a key role in order to
introduce nontrivial parameters. Suppose in fact that, instead of
giving rise to a soft breaking, the term $S_{\sigma }(\underline{\sigma },%
\underline{\alpha },\underline{\beta },\phi )$ would be left invariant by
the $BRST$ operator, \textit{i.e.}
\begin{equation}
sS_{\sigma }(\underline{\sigma },\underline{\alpha },\underline{\beta },\phi
)=0\;.  \label{a13}
\end{equation}
Therefore, owing to the doublet structure of $(\underline{\alpha },%
\underline{\beta })$, a local functional $\widehat{S}_{\sigma }$ should
exist such that
\begin{equation}
S_{\sigma }(\underline{\sigma },\underline{\alpha },\underline{\beta },\phi
)=s\widehat{S}_{\sigma }\;,  \label{a14}
\end{equation}
from which it would follow that
\begin{equation}
\frac{\partial S_{\sigma }}{\partial \underline{\sigma }}=s\frac{\partial
\widehat{S}_{\sigma }}{\partial \underline{\sigma }}\;,  \label{a15}
\end{equation}
which would imply that $\left( \underline{\sigma }\right) $ would be
unphysical parameters. As a consequence, the correlation functions of the
observables of the theory would be independent from $\left( \underline{%
\sigma }\right) $, meaning that the introduction of these parameters would
have no effects on the theory.

\section{The example of the Gribov-Zwanziger Lagrangian}

As a first example, we give a derivation of a refined version of
the Gribov-Zwanziger Lagrangian, see
\cite{Dudal:2007cw,Dudal:2008sp}, within the previous set up. We
start with the Yang--Mills action, as given in eq.$\left(
\ref{a3}\right)$, and we add a set of fields
$\left( \overline{\varphi }_{\mu }^{ac},\varphi _{\mu }^{ac}\right)$, $%
\left( \overline{\omega }_{\mu }^{ac},\omega _{\mu }^{ac}\right) $
transforming as $BRST$ doublets. For the $BRST$ invariant action
we write
\begin{equation}
S_{0}=S_{YM}+S_{gf}\,+s\int d^{4}x\;\left( \overline{\omega }_{\mu
}^{ac}\partial _{\nu }\left( \partial _{\nu }\varphi _{\mu
}^{ac}+gf^{abm}A_{\nu }^{b}\varphi _{\mu }^{mc}\right) -\mu ^{2}\overline{%
\omega }_{\mu }^{ac}\varphi _{\mu }^{ac}\right) \;,  \label{d1}
\end{equation}
with
\begin{eqnarray}
sA_{\mu }^{a} &=&-\left( D_{\mu }c\right) ^{a}\;,  \nonumber \\
sc^{a} &=&\frac{1}{2}gf^{abc}c^{b}c^{c}\;,  \nonumber \\
s\overline{c}^{a} &=&ib^{a}\;,\;\;\;\;\;\,\,sb^{a}=0\;,  \label{d2}
\end{eqnarray}
and
\begin{eqnarray}
s\varphi _{\mu }^{ac} &=&\omega _{\mu }^{ac}\;,\;\;\;\;\;s\omega _{\mu
}^{ac}=0\;,  \nonumber \\
s\overline{\omega }_{\mu }^{ac} &=&\overline{\varphi }_{\mu
}^{ac}\;,\;\;\;\;\;s\overline{\varphi }_{\mu }^{ac}=0\;,\;  \label{d3}
\end{eqnarray}
\begin{equation}
sS_{0}\;=0\;.  \label{d4}
\end{equation}
For the explicit expression of $S_{0}$ one gets
\begin{eqnarray}
S_{0} &=&S_{\mathrm{YM}}+\int d^{4}x\;\left( ib^{a}\partial _{\mu }A_{\mu
}^{a}+\overline{c}^{a}\partial _{\mu }\left( D_{\mu }c\right) ^{a}\right) \;
\nonumber \\
&+&\int d^{4}x\left( \overline{\varphi }_{\mu }^{ac}\partial _{\nu }\left(
\partial _{\nu }\varphi _{\mu }^{ac}+gf^{abm}A_{\nu }^{b}\varphi _{\mu
}^{mc}\right) -\overline{\omega }_{\mu }^{ac}\partial _{\nu }\left( \partial
_{\nu }\omega _{\mu }^{ac}+gf^{abm}A_{\nu }^{b}\omega _{\mu }^{mc}\right)
\right)   \nonumber \\
&-&\int d^{4}x\;\left( g\left( \partial _{\nu }\overline{\omega }_{\mu
}^{ac}\right) f^{abm}\left( D_{\nu }c\right) ^{b}\varphi _{\mu }^{mc}+\mu
^{2}\left( \overline{\varphi }_{\mu }^{ac}\varphi _{\mu }^{ac}-\overline{%
\omega }_{\mu }^{ac}\omega _{\mu }^{ac}\right) \right) \;.
\label{d5}
\end{eqnarray}
The fields $\left( \overline{\varphi }_{\mu }^{ac},\varphi _{\mu
}^{ac}\right) $ are a pair of complex conjugate bosonic fields.
They have dimension one and zero ghost number. Each field has
$4\left( N^{2}-1\right) ^{2}$ components. Similarly, the fields
$\left( \overline{\omega }_{\mu
}^{ac},\omega _{\mu }^{ac}\right) $ are anticommuting, having ghost number $%
(-1,1)$, respectively.
\newline
\newline
From the $BRST\;$doublet structure of the additional fields
$\left( \overline{\varphi }_{\mu }^{ac},\varphi _{\mu
}^{ac},\overline{\omega }_{\mu }^{ac},\omega _{\mu }^{ac}\right) $
it follows that the action $S_{0}$ is equivalent to pure
Yang--Mills theory, leading to the same ultraviolet behavior. The
doublet structure implies in fact that there is an exact
compensation in the Feynman diagrams among the bosonic sector
$\left( \overline{\varphi }_{\mu }^{ac},\varphi _{\mu
}^{ac}\right) $ and the anticommuting one $\left( \overline{\omega
}_{\mu }^{ac},\omega _{\mu }^{ac}\right) $. Also, it is easily
verified that the integration over the auxiliary fields $\left(
\overline{\varphi }_{\mu }^{ac},\varphi _{\mu
}^{ac},\overline{\omega }_{\mu }^{ac},\omega _{\mu }^{ac}\right) $
amounts to introduce a unity in the partition function, meaning
that the physical content of the theory is precisely the same as
that of the Faddeev-Popov action.
 \\\\We   now
introduce a soft breaking of the $BRST$ symmetry in a way which
modifies the low energy sector, while keeping the original
ultraviolet behavior as well as the renormalizability of the
theory, as explained in the last section.
We thus introduce a quadratic term describing the coupling among
the gauge field $A_{\mu }^{a}$ and the bosonic fields $\left(
\overline{\varphi }_{\mu }^{ac},\varphi _{\mu }^{ac}\right) $.
This term will affect the gluon propagator $\langle A_{\mu
}^{a}(k)A_{\nu }^{b}(-k)\rangle $ in the infrared. As a
consequence, the correlation functions of the gauge invariant
quantities, like e.g. $\left\langle F^{2}(x)F^{2}(y)\right\rangle
$, will get modified in the low energy region. For this quadratic
term, one can write the following expression
\begin{equation}
S_{\gamma }=-\gamma ^{2}g\int d^{4}x\;\left( f^{abc}A_{\mu
}^{a}\varphi _{\mu }^{bc}+f^{abc}A_{\mu }^{a}\overline{\varphi
}_{\mu }^{bc}\right)  \;.  \label{d6}
\end{equation}
where $\gamma $ is a parameter with mass dimension one. With the
introduction of the  term $S_{\gamma }$ the action
\begin{equation}
S=S_{0}+S_{\gamma }  \label{d7}
\end{equation}
does not display anymore exact $BRST$ invariance. Instead, one has the
softly broken identity.
\begin{equation}
sS=\gamma ^{2}\Delta _{\gamma }\;,  \label{d8}
\end{equation}
where the soft breaking term is given by
\begin{equation}
\Delta _{\gamma }=\int d^{4}x\,gf^{abc}\left( \left( D_{\mu
}^{am}c^{m}\right) \left( \varphi _{\mu }^{bc}\,+\overline{\varphi }_{\mu
}^{bc}\right) -A_{\mu }^{a}\omega _{\mu }^{bc}\right) \;.  \label{d9}
\end{equation}
Looking at the gluon propagator $\langle A_{\mu }^{a}(k)A_{\nu
}^{b}(-k)\rangle $, one finds
\begin{equation}
\langle A_{\mu }^{a}(k)A_{\nu }^{b}(-k)\rangle =\delta ^{ab}\frac{k^{2}+\mu
^{2}}{k^{4}+\mu ^{2}k^{2}+\gamma ^{4}}\;\left( \delta _{\mu \nu }-\frac{%
k_{\mu }k_{\nu }}{k^{2}}\right) \; \label{d10}
\end{equation}
or
\begin{equation}
\langle A_{\mu }^{a}(k)A_{\nu }^{b}(-k)\rangle =\delta ^{ab}\frac{    1  }{k^{2}+
{\cal M}^2( k^{2})   }\;\left( \delta _{\mu \nu }-\frac{%
k_{\mu }k_{\nu }}{k^{2}}\right) \;  \label{d10b}
\end{equation}
with
\begin{equation}
{\cal M}^2( k^{2}) = \frac{    \gamma ^{4}  }{k^{2}+
  \mu^{2}    }  \;.  \label{d10bb}
\end{equation}
We see therefore that, while keeping the usual ultraviolet
behavior $\simeq 1/k^{2}$ at very high momenta, expression
(\ref{d10}) turns out to be deeply modified in the infrared region
$k\approx 0$ by the presence of the soft parameters $\left( \gamma
,\mu \right) $. The propagator (\ref{d10}) is not of the Gribov
type, as it does not vanish at the origin, thanks to the presence
of the parameter $\mu$ \cite{Dudal:2007cw,Dudal:2008sp}.
Remarkably, this propagator is in good agreement with the most
recent lattice numerical simulations
\cite{Cucchieri:2007md,Bogolubsky:2007ud,Cucchieri:2007rg,Cucchieri:2008fc}.
In particular, it exhibits violation of reflection positivity,  a
feature which is usually interpreted a signal for gluon
confinement. \\\\Without entering into details, it is worth
mentioning here that the action $S $ of expression (\ref{d7}) has
a profound geometrical meaning
\cite{Zwanziger:1989mf,Zwanziger:1992qr}, as it allows to restrict
the domain of integration in the path integral to the so called
Gribov region $\Omega $, defined as the set of gauge fields
fulfilling the Landau condition and for which the Faddeev-Popov
operator $\mathcal{M}^{ab}=$ $-\partial _{\mu }D_{\mu }^{ab}$ is
strictly positive, namely
\begin{equation}
\Omega =\left\{ A_{\mu }^{a};\;\partial _{\mu }A_{\mu }^{a}=0,\;\;\;\mathcal{%
M}^{ab}=-\partial _{\mu }D_{\mu }^{ab}>0\right\} \;.  \label{d11}
\end{equation}
The restriction to this region is needed in order to properly take
into account the existence of the Gribov copies. In particular,
the parameter $\gamma $, known as the Gribov parameter, is not a
free parameter of the theory, being determined in a
self-consistent way through the gap equation
\cite{Zwanziger:1989mf,Zwanziger:1992qr}
\begin{equation}
\frac{\delta \Gamma }{\delta \gamma ^{2}}=0\;,  \label{d12}
\end{equation}
where $\Gamma$ stands for the $1PI$ generating functional. This
equation enables us to express it in terms of the coupling
constant $g$ and of the non-perturbative invariant scale $\Lambda
_{QCD}$. In much the same way, the parameter $\mu $ has been
determined by means of a variational principle, see
refs.\cite{Dudal:2007cw,Dudal:2008sp} for the numerical
characterization of $\left( \gamma ,\mu \right) $. It is worth
mentioning that the introduction of the $BRST$ invariant mass term
in expression  (\ref {d5})
\begin{equation}
\int d^{4}x\;\mu ^{2}\left( \overline{\varphi }_{\mu }^{ac}\varphi _{\mu
}^{ac}-\overline{\omega }_{\mu }^{ac}\omega _{\mu }^{ac}\right) \;.
\end{equation}
follows from the observation that the dimension two condensate $\left\langle
\overline{\varphi }_{\mu }^{ac}\varphi _{\mu }^{ac}-\overline{\omega }_{\mu
}^{ac}\omega _{\mu }^{ac}\right\rangle $ has a non-vanishing value. In fact,
as shown in \cite{Dudal:2007cw,Dudal:2008sp}, one finds
\begin{eqnarray}
\left\langle \overline{\varphi }_{\mu }^{ac}\varphi _{\mu }^{ac}-\overline{%
\omega }_{\mu }^{ac}\omega _{\mu }^{ac}\right\rangle  &=&\frac{3(N^{2}-1)}{%
64\pi }\lambda   \nonumber \\
\lambda ^{4} &=&2g^{2}N\gamma ^{4}\;,
\end{eqnarray}
from which one sees that $\left\langle \overline{\varphi }_{\mu
}^{ac}\varphi _{\mu }^{ac}-\overline{\omega }_{\mu }^{ac}\omega
_{\mu }^{ac}\right\rangle $ is nonzero for non-vanishing Gribov
parameter $\gamma $. Let us also mention that the appearance of
the $BRST$ soft breaking term $\Delta_{\gamma}$ can be traced back
to the properties of the Gribov region. It arises as a consequence
of the fact that any infinitesimal gauge transformation of a field
configuration belonging to $\Omega$ gives rise to a configuration
which lies outside of $\Omega $ \cite{Dudal:2008sp}.\\\\A
remarkable aspect of the action $\left( S_{0}+S_{\gamma }\right) $
is its multiplicative renormalizability\footnote{We remind here
that vacuum terms of the kind $\int d^4x 4(N^2-1)\gamma^4$ have to
be properly taken into account when solving the gap equation
(\ref{d12}). These terms show up in fact during the
renormalization procedure, as it follows by employing the method
of the external sources, as outlined in
\cite{Zwanziger:1989mf,Zwanziger:1992qr,Maggiore:1993wq,Dudal:2007cw,Dudal:2008sp}.}
\cite{Zwanziger:1989mf,Zwanziger:1992qr,Maggiore:1993wq,Dudal:2007cw,Dudal:2008sp}.
Only two independent renormalization factors are needed to account
for all ultraviolet divergences. It should be noticed that both
the Gribov parameter $\gamma $ and the dynamical mass $\mu $ do
not renormalize independently, the corresponding renormalization
factors being given by suitable products of the two independent
renormalization factors of the theory which can be chosen as being
the renormalization of the coupling constant $Z_{g}$ and of the
gluon field $Z_{A}$. The soft breaking $\Delta _{\gamma }$ does
not invalidate the existence of the Slavnov--Taylor identities for
the
renormalized $1PI$ generating functional $%
\Gamma $, namely
\begin{equation}
\mathcal{S}(\Gamma )=\left[ \Delta _{\gamma }\cdot \Gamma \right] \;,
\label{sti}
\end{equation}
where $\left[ \Delta _{\gamma }\cdot \Gamma \right] $ is the generator of
the $1PI$ Green functions with the insertion of the breaking $\Delta
_{\gamma }$ and
\begin{equation}
\mathcal{S}(\Gamma )=\int d^{4}x\left( \frac{\delta \Gamma }{\delta A_{\mu
}^{a}}\frac{\delta \Gamma }{\delta \Omega _{\mu }^{a}}+\frac{\delta \Gamma }{%
\delta c^{a}}\frac{\delta \Gamma }{\delta L^{a}}+b^{a}\frac{\delta \Gamma }{%
\delta \overline{c}^{a}}+\omega _{\mu }^{ac}\frac{\delta \Gamma }{\delta
\varphi _{\mu }^{ac}}+\overline{\varphi }_{\mu }^{ac}\frac{\delta \Gamma }{%
\delta \overline{\omega }_{\mu }^{ac}}\right)   \label{sto}
\end{equation}
with $\Omega _{\mu }^{a}$ and $L^{a}$ being the external sources
coupled to the nonlinear $BRST\;$ transformation of the gluon
field $A_{\mu }^{a}$ and of the ghost field $c^{a}$. Equation
(\ref{sti}) is the quantum generalization of the broken identity
(\ref{d8}). It is worth emphasizing that the Slavnov--Taylor
identities (\ref{sti}) have in fact a powerful predictive
character. They allow us to establish the relationships among the
various Green functions of the theory in a way which takes into
account the presence of the Gribov horizon \cite{Dudal:2008sp}.
\\\\Let us also mention that, due to the presence of the soft
breaking, the mass parameter $\mu^2$ acquires the meaning of a
physical parameter of the theory, even if it has been introduced
through an $s$-exact term, eq.(\ref{d1}). This important feature
has a simple understanding by noticing that, in the absence of the
Gribov parameter, {\it i.e.} $\gamma=0$, the contributions of the
two $BRST$ doublets, $\left( \overline{\varphi }_{\mu
}^{ac},\varphi _{\mu }^{ac}\right) $ and $\left( \overline{\omega
}_{\mu }^{ac},\omega _{\mu }^{ac}\right) $, compensate each other
in an exact way. In fact, setting $\gamma=0$, one immediately
realizes that the gluon propagator in eq.(\ref{d10}) gets
independent from the parameter $\mu^2$. However, in the presence
of the breaking, {\it i.e.} for $\gamma\neq 0$, the compensation
among the two sectors $\left( \overline{\varphi }_{\mu
}^{ac},\varphi _{\mu }^{ac}\right) $ and $\left( \overline{\omega
}_{\mu }^{ac},\omega _{\mu }^{ac}\right)$ does not hold anymore,
due to the fact that the  fields $\left( \overline{\varphi }_{\mu
}^{ac},\varphi _{\mu }^{ac}\right) $ have been coupled to the
gluon field through the quadratic term of eq.(\ref{d6}). As a
consequence, the parameter $\mu^2$ enters in a nontrivial way the
gluon propagator as well as the correlation functions of the
theory. A more formal proof of this property can be given by
making use of the renormalized Slavnov-Taylor identities
(\ref{sti}).

\section{A model for the dynamical quark mass generation}

As second example, we present a possible mechanism for introducing
a dynamical quark mass. To some extent, this is a way to push
further the gate opened by Gribov-Zwanziger in the gluon sector,
using the idea that all partons of a confining and  asymptotically
free theory should get a propagator with a momentum dependence as
in eq.(\ref{d10}), since they all have very analogous physical
status. The mechanism will consist in introducing $BRST$~doublets
in correspondence with   the quarks, and check the possible
$BRST$~symmetry breaking soft terms that are allowed by power
counting. \\\\We thus supplement the Yang--Mills action of
expression $\left( \ref{a3}\right) $ by the fermionic matter term
\begin{equation}
S_{\psi }=\int d^{4}x\;\overline{\psi }\gamma _{\mu }D_{\mu }\psi \;,
\label{c}
\end{equation}
with
\begin{eqnarray}
s\psi &=&c^{a}T^{a}\psi \;,  \nonumber \\
sS_{\psi } &=&0\;.  \label{c1}
\end{eqnarray}
Further, we introduce two $BRST$ doublets of spinor fields, $\left( \varphi
_{F},\omega _{F}\right) $ and $\left( \lambda _{F},\eta _{F}\right) $
transforming as
\begin{eqnarray}
s\varphi _{F} &=&\omega _{F}\;,\;\;\;\;\;\;s\omega _{F}=0\;,  \nonumber \\
s\eta_{F} &=&\lambda_{F}\;,\,\,\,\,\,\,\,\,\,\,s\lambda_{F}=0\;.
\label{c3}
\end{eqnarray}
The fields $\left( \varphi _{F},\omega _{F}\right) $ have
dimension $1/2$ and ghost number $\left( 0,1\right) $,
respectively, while $\left( \lambda _{F},\eta _{F}\right) $ have
dimension $3/2$ and ghost number $\left( 0,-1\right) $.
\\\\According to the set-up of Sect.2, the invariant action
describing the propagation of these fields is obtained from an
exact $BRST\;$variation, namely
\begin{eqnarray}
S_{\varphi \lambda } &=&s\int d^{4}x\left( -\overline{\eta }_{F}\partial
^{2}\varphi _{F}+\overline{\varphi }_{F}\partial ^{2}\eta _{F}+m^{2}(%
\overline{\eta }_{F}\varphi _{F}-\overline{\varphi }_{F}\eta _{F})\right)
\nonumber \\
&=&\int d^{4}x\left( -\overline{\lambda }_{F}\partial ^{2}\varphi
_{F} - \overline{\varphi }_{F}\partial ^{2}\lambda
_{F}-\overline{\eta }_{F}\partial ^{2}\omega _{F}+\overline{\omega
}_{F}\partial ^{2}\eta
_{F}\right.  \nonumber \\
&&\;\;\;\;\;\;\;\left. +m^{2}(\overline{\lambda }_{F}\varphi _{F}+\overline{%
\varphi }_{F}\lambda _{F}+\overline{\eta }_{F}\omega _{F}-\overline{\omega }%
_{F}\eta _{F})\right) \;.  \label{c4}
\end{eqnarray}
For the soft breaking term of the $BRST$ symmetry we get
\begin{equation}
S_{M}=M_{1}^{2}\int d^{4}x\left( \overline{\varphi }_{F}\psi +\overline{\psi
}\varphi _{F}\right) -M_{2}\int d^{4}x\left( \overline{\lambda }_{F}\psi +%
\overline{\psi }\lambda _{F}\right) \;.  \label{c5}
\end{equation}
From expression $\left( \ref{c5}\right) $ one realizes that, in a way
similar to the case of the Gribov-Zwanziger action, the additional fields $%
\varphi _{F}$, $\lambda _{F}$ couple linearly to the matter field
$\psi
$. As a consequence, the propagator $\left\langle \psi (k)\overline{\psi }%
(-k)\right\rangle $ of the matter field will get deeply modified
in the infrared by the presence of the soft mass parameters
$m,M_{1},M_{2}$. In fact, one finds
\begin{equation}
\left\langle \psi (k)\overline{\psi }(-k)\right\rangle =\frac{i\gamma _{\mu
}k_{\mu }+\mathcal{A}(k)}{k^{2}+\mathcal{A}^{2}(k)}\;,  \label{c6}
\end{equation}
where the mass function $\mathcal{A}(k)$ is given by
\begin{equation}
\mathcal{A}(k)=\frac{2M_{1}^2 M_{2}}{k^{2}+m^{2}}\;.  \label{c7}
\end{equation}
The function $\mathcal{A}(k)  $  is completely analogous to the
function ${\cal M}^2(k^2)$,  which has appeared in  the propagator
for the gluon in {eq.(\ref{d10b}). Its geometrical interpretation
is however lacking, since the Gribov phenomenon is left untouched
by the quark dependance. For the quarks, the demand of a
modification of the propagator is thus motivated by their
confinement and by the chiral symmetry breaking, and it is made
possible in quantum field theory by this general mechanism
involving the introduction of soft breaking terms using doublets.
\\\\Interestingly, expression $\left( \ref{c7}\right) $ provides a
good fit for the dynamical quark mass in the infrared region, as
reported by the lattice numerical simulations of the quark
propagator in the Landau gauge, see for instance
\cite{Parappilly:2005ei,Furui:2006ks}.
\\\\Although being out of the main goal of the present letter, let
us mention that the renormalizability of the spinor action $\left(
S_{\varphi \lambda }+S_{M}\right) $ stems from the fact that the
fields $\left( \varphi _{F},\lambda _{F}\right) $ , $\left( \omega
_{F},\eta _{F}\right) $ are introduced in a way which enables us
to write a local set of linearly broken identities, as expressed
by
\begin{equation}
\frac{\delta \left( S_{\varphi \lambda }+S_{M}\right) }{\delta \overline{%
\eta }_{F}}=-\partial ^{2}\omega _{F}+m^{2}\omega _{F}\;,  \label{c8}
\end{equation}
\begin{equation}
\frac{\delta \left( S_{\varphi \lambda }+S_{M}\right) }{\delta \overline{%
\omega }_{F}}=\partial ^{2}\eta _{F}-m^{2}\eta _{F}\;,  \label{c9}
\end{equation}
\begin{equation}
\frac{\delta \left( S_{\varphi \lambda }+S_{M}\right) }{\delta \overline{%
\lambda }_{F}}=-\partial ^{2}\varphi _{F}+m^{2}\varphi _{F}-M_{2}\psi \;,
\label{c10}
\end{equation}
\begin{equation}
\frac{\delta \left( S_{\varphi \lambda }+S_{M}\right) }{\delta \overline{%
\varphi }_{F}}=-\partial ^{2}\lambda _{F}+m^{2}\lambda _{F}+M_{1}^{2}\psi \;.
\label{c11}
\end{equation}
Thanks to the fact that the right hand sides of eqs.$\left(
\ref{c8}\right)-\left( \ref{c11}\right) $ is linear in the fields,
it turns out that these equations can be converted in a set of
powerful Ward identities which constrain very much the possible
counterterms which can show up at quantum level, ensuring that the
action $\left( S_{\varphi \lambda }+S_{M}\right) $ is stable
against quantum corrections. The reader might have in fact
remarked that equations $\left( \ref{c8}\right) -\left(
\ref{c11}\right) $
follow from the fact that the quadratic terms in expression $\left( \ref{c4}%
\right) $ contain simple space-time derivatives, e.g. $\overline{\eta }%
_{F}\partial ^{2}\varphi _{F}$, and not covariant derivatives of the kind $%
\overline{\eta }_{F}D^{2}\varphi _{F}$. This term would jeopardize the
linearity in the quantum fields of eqs.$\left( \ref{c8}\right) -\left( \ref
{c11}\right) $, as quadratic and cubic terms in the fields would be present.
These terms have to be treated as composite operators, requiring a proper
renormalization procedure. In summary, eqs.$\left( \ref{c8}\right) -\left(
\ref{c11}\right) $ would loose their usefulness. Moreover, as much as in the
case of a $\phi ^{4}-$theory, the presence of a term of the kind $\overline{%
\eta }_{F}D^{2}\varphi _{F}$ would require the introduction of a
quartic spinor interaction like $\rho \left( \overline{\eta
}_{F}\varphi _{F}\right) \left( \overline{\varphi }_{F}\eta
_{F}\right) $ with its own dimensionless parameter $\rho$,
implying that we would be leaving the space of soft parameters.
Let us finally mention that a similar set of Ward identities can
be derived in the case of the Gribov-Zwanziger action
\cite{Zwanziger:1989mf,Zwanziger:1992qr,Maggiore:1993wq,Dudal:2007cw,Dudal:2008sp}.

\section{Conclusion}
In this letter we have illustrated a proposal in order to
introduce non-perturbative infrared effects leading to a
modification of the long distance behavior of the correlation
functions of the theory. \\\\The whole idea relies on the
introduction of additional fields giving rise to $BRST$ doublets.
These fields do not modify the set of original observables
of the theory, identified here with the cohomology classes of the
$BRST$ operator. Subsequently, a set of soft parameters are
introduced through terms which give rise to a soft breaking of the
$BRST$ invariance. The simplest way to do this is by demanding
that the doublet fields couple linearly to the original fields of
the theory, so that the soft breaking is given by terms quadratic
in the fields. As a consequence, the propagators of the original
fields of the theory get deeply modified in the infrared. In turn,
the correlation functions of the local observables of the theory
gets modified too in the low energy region. \\\\We have pointed
out that such kind of coupling enables us to write down suitably
linearly broken Ward identities which can ensure that the theory
remains renormalizable in the presence of the breaking. Moreover,
we have seen that the presence of this breaking allows us to give
an elementary algebraic proof of the fact that the corresponding
soft parameters are physical parameters and not unphysical ones.
\\\\The Gribov-Zwanziger action implementing the restriction in
the Feynman path integral to the Gribov region $\Omega$ has been
revised within the present set up. Finally, we have proposed a
model which might be of a certain interest in the study of the
dynamical quark mass generation. The details of the
renormalizability of this model as well as a discussion of
possible gap equations allowing to give an estimate of the
parameters $m, M_1,M_2$ will be the object of a more extended work
\cite{prep}.

\section*{Acknowledgments}
We thank Daniel Zwanziger for useful comments and discussions. The
Conselho Nacional de Desenvolvimento Cient\'{i}fico e
Tecnol\'{o}gico (CNPq-Brazil), the Faperj, Funda{\c{c}}{\~{a}}o de
Amparo {\`{a}} Pesquisa do Estado do Rio de Janeiro, the SR2-UERJ
and the Coordena{\c{c}}{\~{a}}o de Aperfei{\c{c}}oamento de
Pessoal de N{\'{i}}vel Superior (CAPES) are gratefully
acknowledged for financial support. This work has been also
partially supported by the contract ANR (CNRS-USAR),
\texttt{05-BLAN-0079-01}.

\end{document}